\documentclass[twocolumn,tighten]{aastex61}
\usepackage{amssymb, amsfonts, amsmath,framed}

\usepackage{bm}
\expandafter\ifx\csname package@font\endcsname\relax\else
 \expandafter\expandafter
 \expandafter\usepackage
 \expandafter\expandafter
 \expandafter{\csname package@font\endcsname}
\fi
\expandafter\ifx\csname package@font\endcsname\relax\else
 \expandafter\expandafter
 \expandafter\usepackage
 \expandafter\expandafter
 \expandafter{\csname package@font\endcsname}
\fi
\def\bq{\begin{equation}}
\def\eq{\end{equation}}
\def\bqy{\begin{eqnarray}}
\def\eqy{\end{eqnarray}}

\begin{document}

\title{Reduced Diversity of Life Around Proxima Centauri and TRAPPIST-1}

\correspondingauthor{Manasvi Lingam}
\email{manasvi.lingam@cfa.harvard.edu}

\author{Manasvi Lingam}
\affiliation{Harvard-Smithsonian Center for Astrophysics, 60 Garden St, Cambridge, MA 02138, USA}
\affiliation{John A. Paulson School of Engineering and Applied Sciences, Harvard University, 29 Oxford St, Cambridge, MA 02138, USA}

\author{Abraham Loeb}
\affiliation{Harvard-Smithsonian Center for Astrophysics, 60 Garden St, Cambridge, MA 02138, USA}

\begin{abstract}
The recent discovery of potentially habitable exoplanets around Proxima Centauri and TRAPPIST-1 has attracted much attention due to their potential for hosting life. We delineate a simple model that accurately describes the evolution of biological diversity on Earth. Combining this model with constraints on atmospheric erosion and the maximal evolutionary timescale arising from the star's lifetime, we arrive at two striking conclusions: (i) Earth-analogs orbiting low-mass M-dwarfs are unlikely to be inhabited, and (ii) K-dwarfs and some G-type stars are potentially capable of hosting more complex biospheres than the Earth. Hence, future searches for biosignatures may have higher chances of success when targeting planets around K-type stars. 
\end{abstract}

\section{Introduction}
The discovery of thousands of exoplanets by the Kepler satellite \citep{Bor16} has rekindled interest in the fundamental question: ``Are we alone?'' In order to address this question, astronomers have been searching for planets in the habitable zone (HZ), i.e. the region around the host star that is theoretically capable of supporting liquid water. It is now estimated that the Milky Way galaxy hosts $\sim 10^{10}$ exoplanets in the HZ, of which the majority orbit the lowest mass stars, namely M-dwarfs \citep{DC15}. 

It is relatively easier to detect exoplanets in the HZ around low-mass stars, and they also happen to be numerous. Therefore, studies of the habitability of exoplanets around M-dwarfs have become increasingly common in recent times \citep{SBJ16}. This field received a major impetus over the past year with the discovery of Proxima b \citep{AE16} and the TRAPPIST-1 system \citep{Gill17}. The former is the closest exoplanet to Earth, while the latter comprises of at least seven Earth-sized planets, three of which are situated in the HZ. 

However, it is important to recognize that a planet in the HZ is not necessarily habitable. One of the prerequisites for life-as-we-know-it is the existence of an atmosphere. Non-thermal atmospheric losses for HZ exoplanets orbiting M-dwarfs can be quite extensive - a consequence of their close distances to the host star subjecting them to intense stellar winds and extreme space weather conditions \citep{DLMC,Aira17,LiLo17}. In this Letter, we combine constraints imposed by stellar wind-induced atmospheric erosion and the star's finite lifetime with a simple model for the evolution of biodiversity on Earth. Amongst other findings, we demonstrate that lower mass M-dwarfs are not likely to host any lifeforms, and that future searches for life should be directed towards K- and G-type stars.

\section{A simple model for terrestrial biodiversity}
When confronted with intricate concepts such as biological complexity and biodiversity, it is by no means easy to settle upon a universal metric that fully encapsulates these concepts \citep{Mag04}. Biological complexity has often been couched in the language of information theory, but a wide range of alternatives have also been investigated \citep{Ada02}. Similarly, biodiversity, a portmanteau of biological diversity, encompasses variations in species and ecosystems across space and time \citep{Ros95}.

For the purposes of this paper, we shall confine ourselves to simple concepts that can be defined and envisioned in an unambiguous manner. We will focus on evaluating the species richness, i.e. the number of distinct species, on exoplanets. It is reasonable to hypothesize that: (i) larger planets have a higher species diversity because of global species-area power-law relationships with positive exponents \citep{Mag04}, and (ii) older planets have had longer time for speciation to occur, thereby leading to increased species diversity.

Apart from (i) and (ii), one may rightly expect other factors such as fluctuations in the climate (e.g. temperature) and type of habitat (terrestrial or aquatic) to play an important role in influencing species diversity \citep{Ben09}. We restrict our discussion to (ii) because of its significance, and the fact that the age constitutes a readily observable physical parameter for exoplanets; we exclude (i) since the restricted size range of rocky exoplanets \citep{CK17} makes it a sub-dominant contribution \citep{LiLo17}.

Next, we ask: what would constitute an effective `fitting' function for the total species richness as a function of time? We do not consider events such as mass extinctions and the Cambrian explosion \citep{KN17} occurring on exoplanets; these phenomena are ostensibly stochastic, and the consequent extinction/proliferation of species is not predictable \emph{a priori}. Instead, we note that the paradigm of exponential growth has been established to be fairly accurate in modelling the number of species $N$ as a function of time $t$ \citep{PuHe00,Ben09},
\begin{equation} \label{ExpDiv}
    N(t) = \exp\left(\frac{t}{\tau}\right) - 1,
\end{equation}
where $\tau$ can be viewed as a characteristic timescale of species diversification. This function has been chosen such that $N(0) = 0$, and we can multiply the RHS with an additional parameter if necessary. In reality, the exponential growth will eventually saturate due to limited resources, and a logistic function would be more suitable \citep{PuHe00}. There is no definitive evidence that this saturation, at a maximum species diversity $\left(N_\mathrm{max}\right)$, has been realized yet on Earth.

Recent studies have concluded that there exist $\sim 8.7 \times 10^6$ eukaryotic species \citep{MTASW}, and $10^{11}$ to $10^{12}$ microbial taxa \citep{LL16} on Earth indicating that the latter constitute an ``unseen majority'' \citep{WCW98}. There is, however, a considerable degree of uncertainty since some estimates place the total number of microbial species in the millions \citep{SGMET}. Setting $t \sim 4.4$ Gyr and $N(t) \sim 10^{11}$, we find that $\tau \sim 174$ Myr; in contrast, choosing $N(t) \sim 10^7$ leads us to $\tau \sim 274$ Myr. We have chosen $4.4$ Gyr as this represents the time elapsed after the initial formation of oceans and continental crust \citep{WVPG}, thereby enabling evolution.

With the inferred value of $\tau$, we can solve for the time $t_0$ when the first species arose, namely $N\left(t_0\right) = 1$. For the two values of $\tau$, we find $t_0 \sim 121$ Myr and $t_0 \sim 190$ Myr implying that the timescale for abiogenesis could have been $\lesssim 200$ Myr. This model suggests that the first lifeforms appeared $t-t_0$ Gyr in the past, i.e. around $4.2$-$4.3$ Gya. Recently, some preliminary evidence has emerged that might favour the emergence of life on Earth as early as $4.1$-$4.3$ Gya \citep{BBHM,Dod17}. Hence, it is clear that the model (\ref{ExpDiv}) for species richness is surprisingly accurate, given the simplifications involved. 

We can examine a few other events in Earth's geological record by means of the above model. Neglecting the second term on the RHS of (\ref{ExpDiv}), let us suppose that the species richness is $N_1$ and $N_2$ at times $t_1 = t$ and $t_2 = t + \Delta t$ respectively. If we denote the $e$-folding timescale by $\tau_c$, we obtain
\begin{equation}
    \frac{N_2}{N_1} \approx \exp\left(\frac{\Delta t}{\tau_c}\right).
\end{equation}
For the Cambrian explosion, with $\Delta t \sim 25$ Myr \citep{ELT11}, and $N_2/N_1 \sim 10$ \citep{Mar04}, we find $\tau_c \sim 11$ Myr. On comparing this with the global value of $\tau$ determined earlier, we conclude that the rate of diversification was about an order of magnitude higher during the Cambrian period; this conclusion is broadly consistent with results that have been derived from phylogenetic studies \citep{Butt07}.

We reiterate that (\ref{ExpDiv}) is valid only up to some geological timescale as the planet's biosphere will deteriorate due to the runaway greenhouse effect and evaporation of oceans \citep{CK92}, although astrophysical catastrophes are rarer \citep{SBL17,LinLo17}.

\section{Biodiversity and its relation to the host star}

\begin{figure*}
$$
\begin{array}{cc}
  \includegraphics[width=7.5cm]{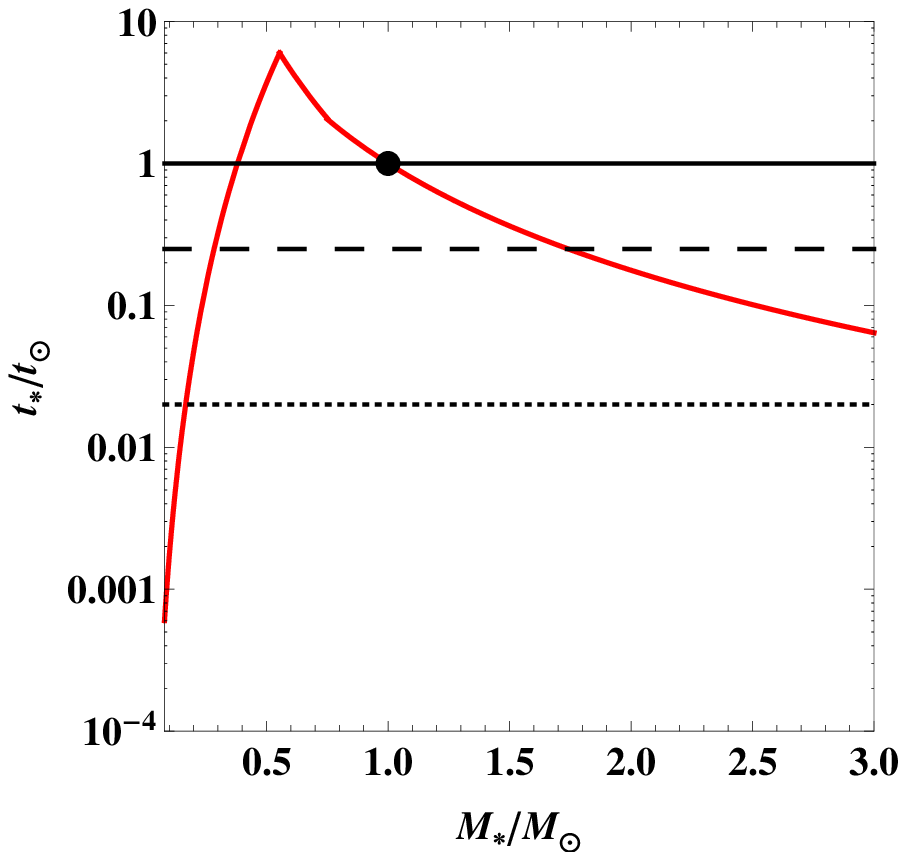} &  \includegraphics[width=7.3cm]{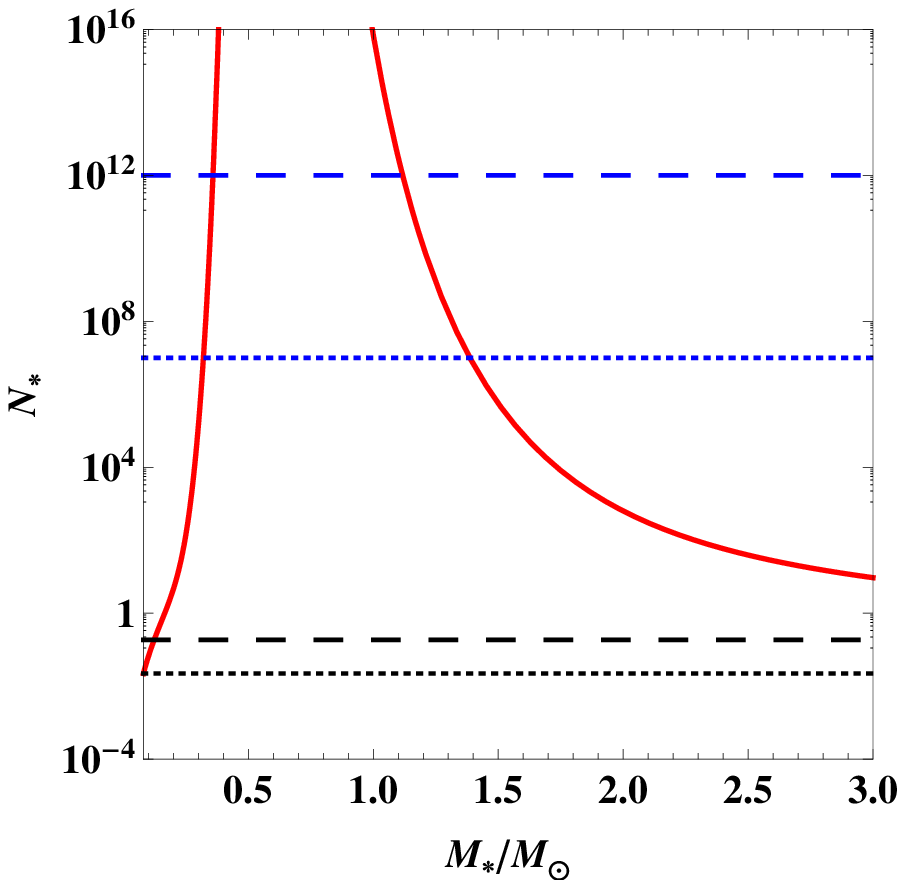}\\
\end{array}
$$
\caption{Left-hand panel: maximum timescale $\left(t_\star\right)$ over which evolution can occur as a function of stellar mass. The dashed and dotted lines correspond to the putative timescales of terrestrial eukaryogenesis and abiogenesis respectively, while the solid line denotes the solar lifetime $\left(t_\odot\right)$. Right-hand panel: peak diversity $\left(N_\star\right)$ attainable as a function of stellar mass. The blue dashed and dotted lines represent the \emph{current} microbial and eukaryotic species diversity on Earth. The black dashed and dotted lines respectively denote the peak values attainable by Earth-analogs orbiting Proxima Centauri and TRAPPIST-1.}
\label{Fig1}
\end{figure*}

Hitherto, we have presented our model (\ref{ExpDiv}) and verified its accuracy in modelling species richness on Earth. We wish to apply it now to exoplanets.

Consider a planet in the HZ of its host star. Naively, we would expect that the stellar lifetime is the maximum period of time that the planet remains in the HZ. Hence, a longer lifetime should translate to a higher species diversity on account of the exponential amplification inherent in (\ref{ExpDiv}) until saturation; after $N = N_\mathrm{max}$ is reached, no further gain is possible. As the stellar lifetime is inversely correlated with its mass, one will therefore expect life to exist on low-mass stars in the cosmic future \citep{LBS16}. However, a crucial factor has recently been recently identified in the context of planetary habitability for M-dwarfs: as these planets are situated close to their host star, they are subject to intense dynamic pressure from the solar wind and strong ultraviolet (UV) radiation, thereby leading to rapid atmospheric stripping by the stellar wind \citep{DLMC,Aira17}. We do not tackle thermal atmospheric escape mechanisms as they are not expected to pose a threat to heavier species on Earth-sized planets \citep{BBMNS,ZC17}.

If the star's mass is higher, the maximum time that the planet can remain in the HZ (the star's overall lifetime) is lowered. However, if the star's mass is lower, the planet's atmosphere can be depleted over short timescales. Thus, from these considerations, we see that a ``Goldilocks zone'' for peak biological diversity emerges naturally as an outcome.

\citet{LiLo17} presented a phenomenological expression for atmospheric erosion by stellar winds, and applied it to an unmagnetized ``Earth-analog'' in the absence of outgassing. The final formula depended on both the star's mass and rotation and, for the sake of simplicity, we set the stellar rotation rate to equal that of the Sun; using the general formula does not alter our results significantly if the rotation rate is comparable to the solar value. The planet's atmosphere is completely depleted over a timescale $t_{SW}$ given by
\begin{equation} \label{LFFixPPv1}
t_{SW} \sim  100 t_\odot\, \left(\frac{M_\star}{M_\odot}\right)^{4.76},
\end{equation}
where $M_\star$ is the stellar mass, $t_\odot \sim 10$ Gyr and $M_\odot$ denote the Sun's total lifetime and mass respectively. For Proxima Centauri, the above formula leads to $t_{SW} \sim 50$ Myr, a value that falls within the range of $10$-$100$ Myr for ion escape losses obtained through detailed numerical simulations \citep{DLMC,Aira17}. We note that (\ref{LFFixPPv1}) would represent an upper bound for the atmospheric loss since atmospheric erosion is enhanced during the active early epoch \citep{Jak17},\footnote{From a theoretical standpoint, it arises as a consequence of
\begin{equation}
\dot{M} \propto \left(\frac{1}{\mathcal{T}_s}\right)^2,
\end{equation}
where $\dot{M}$ is the atmospheric escape rate from stellar wind stripping and $\mathcal{T}_s$ is the age of the star \citep{WMZL,LiLo17}; the formula is valid for solar-type stars. Hence, for weakly magnetized planets in our Solar system, the escape rates were $\sim 100$ times higher $4$ Gya compared to their present-day values.} and (ii) coronal mass ejections (CMEs) are capable of enhancing the escape rates by an order of magnitude \citep{Dong15}.

In contrast, the stellar lifetime $t_\ell$ \citep{LBS16} is expressible as
\begin{eqnarray}
t_\ell &\sim& 1.00\,t_\odot\,\left(\frac{M_\star}{M_\odot}\right)^{-2.5} \quad 0.75 M_\odot < M_\star < 3 M_\odot \\
t_\ell &\sim& 0.76\, t_\odot\, \left(\frac{M_\star}{M_\odot}\right)^{-3.5} \quad 0.25 M_\odot < M_\star \leq 0.75 M_\odot \nonumber \\
t_\ell &\sim& 5.30\, t_\odot\, \left(\frac{M_\star}{M_\odot}\right)^{-2.1} \quad 0.08 M_\odot < M_\star \leq 0.25 M_\odot \nonumber
\end{eqnarray}

Thus, in order to compute the peak species diversity achievable, we must consider the maximum timescale $t_\star$ over which biological growth can occur. From our prior arguments, $t_\star \equiv \mathrm{min}\{t_{SW},t_\ell\}$; the first embodies the constraint set by atmospheric erosion, and the second represents the constraint arising from the age of the host star. In reality, the duration over which the planet is `habitable' is always lesser than the host star's main-sequence lifetime \citep{RCOW}. Moreover, there may exist other relevant timescales that merit consideration; for e.g., as noted earlier, the exponential growth is not operative throughout the planet's lifetime since the species diversity will saturate at some timescale, and then decline.

We may now determine $N_\star \equiv N\left(t_\star\right)$ as a function of $M_\star$. Considering an Earth-analog, we choose $\tau \sim 274$ Myr computed previously since it leads to a value of $t_0$ that is more commensurate with observational evidence; if a different $e$-folding time is adopted, the value of $N$ changes but the essential qualitative features remain.

Fig. \ref{Fig1} shows the maximum timescale over which biological complexity and diversification can occur, as well as the corresponding value of the species diversity. Note that the latter has been truncated after $N\left(t_\odot\right)$ because exponential amplification cannot operate \emph{ad infinitum}; furthermore, this value is already several orders of magnitude higher than the current species diversity on Earth. Based on Fig. \ref{Fig1} and the preceding analysis, the following inferences can be drawn.

\begin{itemize}
    \item The highest value of $t_\star$ is attained at $M_\mathrm{max} = 0.55 M_\odot$ and $t_\mathrm{max} = 6 t_\odot$. These values correspond to a K-dwarf whose lifetime is $\sim 60$ Gyr. 
    \item If we choose a putative timescale of $t_0 \sim 200$ Myr for life to have originated on Earth, host stars with masses $M_\star < 0.17 M_\odot$ are likely to deplete their Earth-analogs of atmospheres prior to this time interval. Although it does not necessarily imply that stars such as Proxima Centauri and TRAPPIST-1 cannot host life-bearing planets, their prospects of doing so appear to be rather minimal; in fact, the probability for these two stars is non-zero only when $t_0 \lesssim 50$ Myr.
    \item Specifying a timescale of $\sim 2.5$ Gyr for the origin of eukaryotes \citep{Knoll14}, only stars that lie within the range $0.28 < M_\star < 1.74 M_\odot$ satisfy the dual criteria of: (i) having sufficiently long lifespans, and (ii) being sufficiently robust against atmospheric erosion. Thus, higher-mass M-dwarfs, all K-type and (most) G-type stars appear to be conducive to the emergence of complex life.
    \item For $0.38 M_\odot < M_\star < M_\odot$, we find $t_\star > t_\odot$, implying that these systems could, in principle, enable evolution (and thus speciation) to occur over a longer period of time than on Earth. Hence, \emph{ceteris paribus}, stars in this mass interval are potentially more capable of hosting biospheres complex than our own as well as facilitating noogenesis. Given that K- and G-type stars span most of this mass range, we suggest that future searches for planetary biosignatures and technosignatures should prioritize these stars as targets, at least insofar putative biospheres with intelligent life are concerned; our conclusion is consistent with previous analyses of this subject \citep{KWR93,HA14,CuGu16}.
    \item The evolutionary lifetime $t_\star$ varies over four orders of magnitude. Hence, the fact that the solar value is less than an order of magnitude compared to the peak $\left(t_\mathrm{max}\right)$ suggests that our existence, orbiting the Sun, is not a coincidence. Instead, it can be explained by the fact that evolution has a longer time interval over which it can function compared to lower mass stars which are more abundant.
    \item As $N_\star$ involves an exponential factor, we see that its value is highly sensitive to $M_\star$. Hence, planets in the HZ around stars lying \emph{outside} the range $0.38 M_\odot < M_\star < M_\odot$ (characterized by $t_\star < t_\odot$) may host biospheres far less diverse than Earth. 
    \item The above fact is likely to be especially true for planets around very low-mass stars such as Proxima Centauri $\left(0.12 M_\odot\right)$ and TRAPPIST-1 $\left(0.08 M_\odot\right)$, indicating that they would not have either simple, complex or intelligent lifeforms.
    \item Based on the extensively documented diversity-stability relationships on Earth, more complex biospheres are likely to promote enhanced stability and ecosystem functionality in some aspects \citep{IC07,Card12}, thereby increasing their chances of survival over long timescales. Hence, a higher value of $t_\star$ corresponds to more time for evolution, and this process should, in turn, beget diverse and long-lived biospheres that are more detectable.
    \item Short-lived biospheres, by definition, also have a lesser duration of time for attaining complexity through vital evolutionary transitions \citep{JMS95,KB00,Sza15}, for e.g., eukaryogenesis, multicellularity, eusociality. Hence, a planet that does not pass through a series of complex evolutionary steps will be incapable of giving rise to intelligence \citep{Cart83,Wat08}.
    \item By choosing a low-mass cutoff of $0.38 M_\odot$ based on the above considerations, and using Fig. 4 of \citet{LBS16}, we conclude that our existence in the current cosmic epoch has a probability of $\sim 1\%$-$10\%$. The new estimate demonstrates that our presence is less anomalous compared to the result of allowing all stars to host life, where the value is $0.1\%$ \citep{LBS16}. 
\end{itemize}
A few important points must be borne in mind with regards to the above results. Our predictions were based on the consideration of only two, albeit important, processes: stellar lifetime and wind-induced atmospheric stripping. Furthermore, the latter entailed three further simplifications: the star's rotation rate is chosen to be comparable to the Sun, while the planet is assumed to have a weak magnetic field and a $1$ bar atmosphere. As a habitability is a multi-dimensional phenomenon, there are several other important factors that must be duly taken into consideration. Our estimates for the timescale for evolution and ensuing biodiversity represent the \emph{maximal} values attainable; in actuality, if the planet's age is sufficiently low, it would not host life currently even though it can do so in the future.

\section{Conclusions}
The ubiquity and diversity of life on Earth is a remarkable feature, especially in light of recent evidence indicating that the planet is home to $\gtrsim 10^{12}$ species. In this paper, we made use of a simple model for species richness over time based on exponential amplification, and showed that it yielded predictions for abiogenesis and the Cambrian explosion on Earth that are broadly consistent with prior studies.

Next, we identified two key processes that limit the enhancement of biodiversity on exoplanets - the first was the finite lifetime of the star, whilst the other was atmospheric erosion by the stellar wind. By incorporating these two factors, we computed the maximum possible timescale $\left(t_\star\right)$ over which species diversity can be enhanced for Earth-analogs. Using this timescale in conjunction with the model of species diversity, we showed that planets orbiting stars within the mass interval $0.38 M_\odot < M_\star < M_\odot$ could potentially host biospheres more complex than our own; the peak is found to occur at $0.55 M_\odot$. In contrast, stars outside the range $0.28 < M_\star < 1.74 M_\odot$ may not have had sufficient time for eukaryotic life to evolve, and those with masses $<0.17 M_\odot$ might lack life altogether (absence of abiogenesis). The putative existence of exponential growth suggests that mass cutoffs between stars hosting planets with complex, primitive and non-existent biospheres would be both sharp and located in a narrow range. 

If these conclusions are valid, they have a number of important implications. The peak value of biodiversity favored by our model $\left(0.55 M_\odot\right)$ is virtually identical to the mass of Kepler-186 $\left(0.54 M_\odot\right)$ known to host a planet, Kepler-186f, in the conservative HZ \citep{Quin14}. Hence, we propose that Kepler-186f should be accorded high priority in future explorations of habitability. Other planets in the HZ worthy of follow-up investigations, as per our criteria, include Kepler-1229b, Kepler-442b and Kepler-62f since their star masses are $0.54 M_\odot$, $0.61 M_\odot$ and $0.69 M_\odot$, respectively. These planets could also possess stronger oceanic tides that play an important role in enhancing the prospects of abiogenesis and complex life \citep{Ling17}.

Our existence around a Sun-like star ought not be written off as mere serendipity since it can be explained by evolution unfolding over a relatively long timescale. Future searches for planetary biosignatures and technosignatures are recommended to focus on K- and G-dwarfs from this \emph{specific} viewpoint, although larger M-dwarfs ought not be ruled out. In contrast, lower mass M-dwarfs like Proxima Centauri and TRAPPIST-1 are unlikely to host life provided that the timescale for the origin of life is similar to that on Earth.\footnote{However, the timescale for abiogenesis on exoplanets is not necessarily comparable to Earth's value \citep{SpTu12}.} Short-lived biospheres with limited species diversity have a number of disadvantages including an inclination towards instability and, in probabilistic terms, lower chances of successfully executing the major evolutionary transitions.

In summary, our work establishes the importance of the host star in regulating the biodiversity of its planets. The star's finite lifetime and stellar wind combine to yield a non-monotonic timescale over which evolution can take place. If species diversification is rapid on Earth-analogs, this timescale might consequently serve as an important cutoff that demarcates planets that: (i) are lifeless, (ii) possess primitive/microbial lifeforms, and (iii) host complex (potentially intelligent) organisms. Some of these results could, in principle, also be applicable to habitable exomoons \citep{HWK14} after taking their distinctions into account.

\acknowledgments
We thank Fred Adams and Chuanfei Dong for their helpful comments concerning the paper. This work was supported in part by a grant from the Breakthrough Prize Foundation for the Starshot Initiative, and by the Institute for Theory and Computation (ITC) at Harvard University. 


\end{document}